\documentclass[conference]{IEEEtran}
\IEEEoverridecommandlockouts

\ifCLASSOPTIONcompsoc
  \usepackage[nocompress]{cite}
\else
  \usepackage{cite}
\fi

\ifCLASSINFOpdf
\else
\fi

\usepackage{textcomp}
\usepackage[most]{tcolorbox}
\usepackage{tcolorbox}
\usepackage[final]{pdfpages}
\usepackage{xcolor}

\usepackage{listings}
\usepackage{enumitem}
\usepackage{amsfonts}
\usepackage{caption}
\usepackage{xspace} %
\usepackage[normalem]{ulem}  %
\useunder{\uline}{\ul}{}  %
\usepackage{booktabs} %
\usepackage{multirow} %
\usepackage{adjustbox}
\usepackage{colortbl}
\usepackage{microtype} 
\definecolor{isabelline}{rgb}{0.96, 0.94, 0.93}

\usepackage{scalerel}
\usepackage{tikz}
\usetikzlibrary{svg.path}

\definecolor{orcidlogocol}{HTML}{A6CE39}
\tikzset{
  orcidlogo/.pic={
    \fill[orcidlogocol] svg{M256,128c0,70.7-57.3,128-128,128C57.3,256,0,198.7,0,128C0,57.3,57.3,0,128,0C198.7,0,256,57.3,256,128z};
    \fill[white] svg{M86.3,186.2H70.9V79.1h15.4v48.4V186.2z}
                 svg{M108.9,79.1h41.6c39.6,0,57,28.3,57,53.6c0,27.5-21.5,53.6-56.8,53.6h-41.8V79.1z M124.3,172.4h24.5c34.9,0,42.9-26.5,42.9-39.7c0-21.5-13.7-39.7-43.7-39.7h-23.7V172.4z}
                 svg{M88.7,56.8c0,5.5-4.5,10.1-10.1,10.1c-5.6,0-10.1-4.6-10.1-10.1c0-5.6,4.5-10.1,10.1-10.1C84.2,46.7,88.7,51.3,88.7,56.8z};
  }
}

\newcommand\orcidicon[1]{\href{https://orcid.org/#1}{\mbox{\scalerel*{
\begin{tikzpicture}[yscale=-1,transform shape]
\pic{orcidlogo};
\end{tikzpicture}
}{|}}}}

\usepackage{hyperref} %

\newtcolorbox{boxB}{
    fontupper = \bf\color{main}\footnotesize, %
    boxrule = 0.5pt,
    colframe = main,
    rounded corners,
    arc = 5pt   %
}

\newtcolorbox{boxD}{
    fontupper = \small, 
    colback = sub, 
    colframe = main, 
    boxrule = 0pt, 
    toprule = 2pt, %
    bottomrule = 2pt %
}

\newtcolorbox{boxH}{
    fontupper = \small, 
    colback = sub, 
    colframe = main, 
    boxrule = 0pt, 
    leftrule = 6pt %
}

\newtcolorbox{boxG}{
    enhanced,
    boxrule = 0pt,
    colback = sub,
    borderline west = {1pt}{0pt}{main}, 
    borderline west = {0.75pt}{2pt}{main}, 
    borderline east = {1pt}{0pt}{main}, 
    borderline east = {0.75pt}{2pt}{main}
}    

\newtcolorbox{boxK}{
    fontupper = \small,
    sharpish corners, %
    boxrule = 0pt,
    toprule = 1.0pt, %
    enhanced,
    fuzzy shadow = {0pt}{-2pt}{-0.5pt}{0.5pt}{black!35} %
}

\newcommand*\circled[1]{\tikz[baseline=(char.base)]{
            \node[shape=circle,draw,inner sep=0.5pt] (char) {#1};}}

\newboolean{changes}

\setboolean{changes}{true}
\ifthenelse{\boolean{changes}}
{
}

\newcommand{\ie}{\textit{i.e.,}\xspace}
\newcommand{\eg}{\textit{e.g.,}\xspace}

\newcommand{\etal}{et al.\xspace}

\newcommand{\github}{GitHub\xspace}
\newcommand{\iot}{IoT\xspace}
\newcommand{\openhab}{\textit{OpenHAB}\xspace}
\newcommand{\homeassistant}{\textit{HomeAssistant}\xspace}

\newcommand{\addon}{\textit{add-ons}\xspace}
\newcommand{\addons}{\textit{add-ons}\xspace}
\newcommand{\fm}{\textit{fm}\xspace}
\newcommand{\ftm}{\textit{ftm}\xspace}

\usepackage{pifont}%
\newcommand{\xmark}{\ding{55}}
\newcommand{\cmark}{\ding{51}}%

\newcounter{rcounter}

\newcounter{fcounter}
\newcommand{\finding}[1]{\refstepcounter{fcounter}
\vspace{0.5em}\noindent\fbox{%
  \parbox{0.95\linewidth}{%
 \vspace{0.3em}{\bf
 {Finding~\arabic{fcounter}~(\fnumber{\arabic{fcounter}})}~--} {#1}
\vspace{0.3em}
 }
}
\vspace{0.5em}
}
\newcommand\fnumber[1]{{$\mathcal{F}_{#1}$}}
\newcommand\pnumber[1]{{\bf P#1}}

\newcommand{\secref}[1]{Sec.~\ref{#1}\xspace}
\newcommand{\figref}[1]{Fig.~\ref{#1}\xspace}
\newcommand{\tabref}[1]{Tab.~\ref{#1}\xspace}

\begin{document}
\title{
Testing Practices, Challenges, and Developer Perspectives in
Open-Source IoT Platforms
}

\makeatletter
\newcommand{\linebreakand}{%
  \end{@IEEEauthorhalign}
  \hfill\mbox{}\par
  \mbox{}\hfill\begin{@IEEEauthorhalign}
}
   
\makeatother

\author{
  \IEEEauthorblockN{1\textsuperscript{st}Daniel Rodriguez-Cardenas\, \orcidicon{0000-0002-3238-1229}\ }
  \IEEEauthorblockA{
    \textit{William \& Mary}\\
    Williamsburg, VA, USA \\
    dhrodriguezcar@wm.edu}
  \and
  \IEEEauthorblockN{2\textsuperscript{nd} Safwat Ali Khan\, \orcidicon{0000-0001-8220-4477}\ }
  \IEEEauthorblockA{
    \textit{George Mason University}\\
    Fairfax, VA, USA \\
    skhan89@gmu.edu}
  \and
  \IEEEauthorblockN{3\textsuperscript{rd} Prianka Mandal\, \orcidicon{0009-0009-1329-9818}\ }
  \IEEEauthorblockA{
    \textit{William \& Mary}\\
    Williamsburg, VA, USA \\
    pmandal@wm.edu}
  \linebreakand %
  \IEEEauthorblockN{4\textsuperscript{th} Adwait Nadkarni\, \orcidicon{0000-0001-6866-4565}\ }
  \IEEEauthorblockA{
    \textit{William \& Mary}\\
    Williamsburg, VA, USA \\
    apnadkarni@wm.edu}
  \and
  \IEEEauthorblockN{5\textsuperscript{th} Kevin Moran\, \orcidicon{0000-0001-9683-5616}\ }
  \IEEEauthorblockA{
    \textit{University of Central Florida}\\
    Orlando, FL, USA \\
    kpmoran@wm.edu}
    \and
  \IEEEauthorblockN{6\textsuperscript{th} Denys Poshyvanyk\, \orcidicon{0000-0002-5626-7586}\ }
  \IEEEauthorblockA{
    \textit{William \& Mary}\\
    Williamsburg, VA, USA \\
    denys@cs.wm.edu}
}

\IEEEtitleabstractindextext{%

\begin{abstract}
As the popularity of Internet of Things (IoT) platforms grows, users gain unprecedented control over their homes, health monitoring, and daily task automation. 
However, the testing of software for these platforms poses significant challenges due to their diverse composition, \eg common smart home platforms are often composed of varied types of devices that use a diverse array of communication protocols, connections to mobile apps, cloud services, as well as integration among various platforms. 
This paper is the first to uncover both the practices and perceptions behind testing in IoT platforms, particularly open-source smart home platforms. 
Our study is composed of two key components. First, we mine and empirically analyze the code and integrations of two highly popular and well maintained open-source IoT platforms, \openhab and \homeassistant. Our analysis involves the identification of functional and related test methods based on the \textit{focal method approach}.
We find that \openhab has only $0.04$ test ratio ($\approx 4K$ focal test methods from  $\approx 76K$ functional methods) in Java files, while \homeassistant exhibits higher test ratio of $0.42$, which reveals a significant dearth of testing. Second, to understand the developers' perspective on testing in IoT, and to explain our empirical observations, we survey 80 open-source developers actively engaged in IoT platform development. 
Our analysis of survey responses reveals a significant focus on automated (unit) testing, and a lack of manual testing, which supports our empirical observations, as well as testing challenges specific to IoT. Together, our empirical analysis and survey yield 10 key findings that uncover the current state of testing in IoT platforms, and reveal key perceptions and challenges.
These findings provide valuable guidance to the research community in navigating the complexities of effectively testing IoT platforms.

\end{abstract}

\begin{IEEEkeywords}
Internet of Things, Software Testing, Maintenance, Unit test, Developer study
\end{IEEEkeywords}}

\maketitle

\IEEEdisplaynontitleabstractindextext

\IEEEpeerreviewmaketitle

\section{Introduction}
\label{sec:introduction}

The Internet of Things (IoT) platforms have proliferated the broader computing landscape in recent years and have ushered in new types of digital interactions. Some estimates project that the number of IoT devices will eclipse nearly 30 billion by 2030~\cite{iot-stats}. IoT devices power increasingly popular smart home ecosystems that aid users in automating daily tasks ranging from controlling lights to implementing home security~\cite{kafle_study_2019, manandhar_helion_2019}. IoT devices have also enabled new types of digital health ecosystems that help users to track their fitness and body~\cite{technogym}, and have even begun to power utilities at the scale of cities~\cite{singh-smart-cities}.  While all of these applications of IoT devices and ecosystems have afforded novel, convenient computing paradigms, the engineering of such systems is not without its challenges.

IoT ecosystems are inherently heterogeneous by nature and comprise smartphones, servers, devices, communication hubs, online services, and end-user applications (see \figref{fig:iot}). The design and engineering of each of the individual components of these ecosystems, alongside their integration, pose significant challenges to the design, implementation, maintenance, testing, and evolution of their software components\cite{makhshari2021}. 

Software testing practices, in particular, are important for IoT systems to ensure that both the functional and security/safety-related requirements are properly met. %
While 
prior studies have examined common bugs and general software development challenges for IoT platforms~\cite{makhshari2021,Corno:2019,Corno:2020}, we know little about the current state of software testing in consumer-oriented IoT platforms such as smart homes. 
That is, the research community has a limited understanding of typical software testing activities, processes, deficiencies, and challenges that engineers currently face while working across all of the typical components of IoT platforms. 

To address this research gap, this paper presents an in-depth study of the testing practices, and perspectives, in open source smart home platforms. Our study has two major components. First, we mine and empirically analyze the testing-related code ($\approx 37K$ test methods, from 12.904 test files) of two of the largest and most active open-source IoT platforms, \openhab~\cite{openhab} and \homeassistant~\cite{homeassistant}. Our choice to study these two popular platforms in-depth was guided by two key study design considerations: {\sf (i)} they represent testing practices across the range of IoT platform components, including, but not limited to, core platform code, device integrations, and integrations for online services, and {\sf (ii)} they are entirely open source, \ie not only is the platform code available, but so are the integrations, which are critical from a testing perspective. Using a rigorous, systematic, open-coding process, we identify the key purposes of testing-related activities and their prevalence across our studied platforms. Moreover, we quantify test coverage by defining the metric of {\em test ratio}, which leverages the notion of the {\em focal method (fm)} (\ie a method with defined parameters, which is to be tested~\cite{tufano_methods2test_2022}) and the {\em focal \underline{test} method (ftm)}, \ie a method to test all or part of an {\em fm)}. Intuitively, the test ratio can be defined as the ratio $ftm/ftm+fm$ (see Section~\ref{sec:bakground} for a detailed overview of {\em fm} and {\em ftm}).

Second, we conducted a survey of 80 open-source developers who work on IoT platforms and asked them about their testing practices, experiences, and perspectives regarding the testing of IoT platforms. 
A systematic, thematic analysis of these survey responses helped us understand the {\em rationale} behind current testing practices and pain points experienced by practitioners, and provided actionable insights.

Our study resulted in \textit{10 key findings} (\fnumber{1} -- \fnumber{10}) that characterize the current landscape of software developers' priorities (\fnumber{2}, \fnumber{8}, \fnumber{9}, \fnumber{10}), preferred approaches and perceptions regarding them (\fnumber{3}, \fnumber{4}), and challenges they face (\fnumber{5}, \fnumber{6}, \fnumber{7}) when testing \iot platforms. Particularly, we find a significant lack of testing of both platform components and integrations (\fnumber{1}), characterized by their computed {\em test ratio}.  
A majority, \ie sensors, control-state, sensor components and integrations with popular \addons such as `Hue', `Nest'\cite{nest}, `SamsungTV', `homekit' \addons fall below the test ratio of 0.5. 
Of \homeassistant's 937 \addons, 610 or 65\% have a test ratio of below 0.5 and an average test ratio of 0.42 (\fnumber{1}). Our analysis also reveals what types of test targets developers prioritize, \eg \openhab reports 16 \addons tests, 12 network tests, and includes rule-based and authentication tests (\fnumber{2}). 

Our analysis of survey responses reveals that developers generally focus on performance, scalability, real-world scenarios, real-time collaboration, and security testing (\fnumber{8}, \fnumber{10}). 
We find that while automated testing is the preferred choice among participants for identifying defects early with enhanced reliability (\fnumber{3}), some developers advocate manual testing for human intuition and adaptability in real-world scenarios (\fnumber{4}). Developers also expressed significant challenges unique to IoT platforms, particularly the validation of compatibility across devices and platforms, debugging of firmware updates, and the challenge of addressing issues in low-power mode (\fnumber{5}-\fnumber{7}). Finally, a few developers express concerns about poor documentation and the absence of organizational standards for testing \iot platforms (\fnumber{9}).

In summary, this paper makes the following contributions:

\begin{itemize}
    \item{An empirical investigation into the purpose of testing related code in the \openhab and \homeassistant open source projects.}
    \item{A survey and thematic analysis of responses from 80 open source developers who work on IoT ecosystems, to investigate developer perspectives and experiences regarding testing practices, tools, processes, and challenges.}
    \item{Ten findings derived from the above two studies that inform important directions for future research.}
    \item{A replication package that contains all of our data, analysis code, and qualitative results to facilitate reproduction and replication of our work~\cite{appendix}.}
\end{itemize}

\section{Background}\label{sec:bakground}

The Internet of Things (\iot) represents the interconnectedness of everyday physical objects or `things' via the internet. These objects are equipped with sensors, gateways, software, hubs, and other technologies, enabling them to collect and exchange data not only among themselves but also with other systems and applications via the Internet. In our study, we outline our research questions concerning the landscape of the most common \iot platforms, practices, and practitioners' perspectives. This section offers an overview of \iot platforms, including their layers, components, and common tests.

\begin{figure}[ht]
  \centering
  \includegraphics[width=0.45\textwidth]{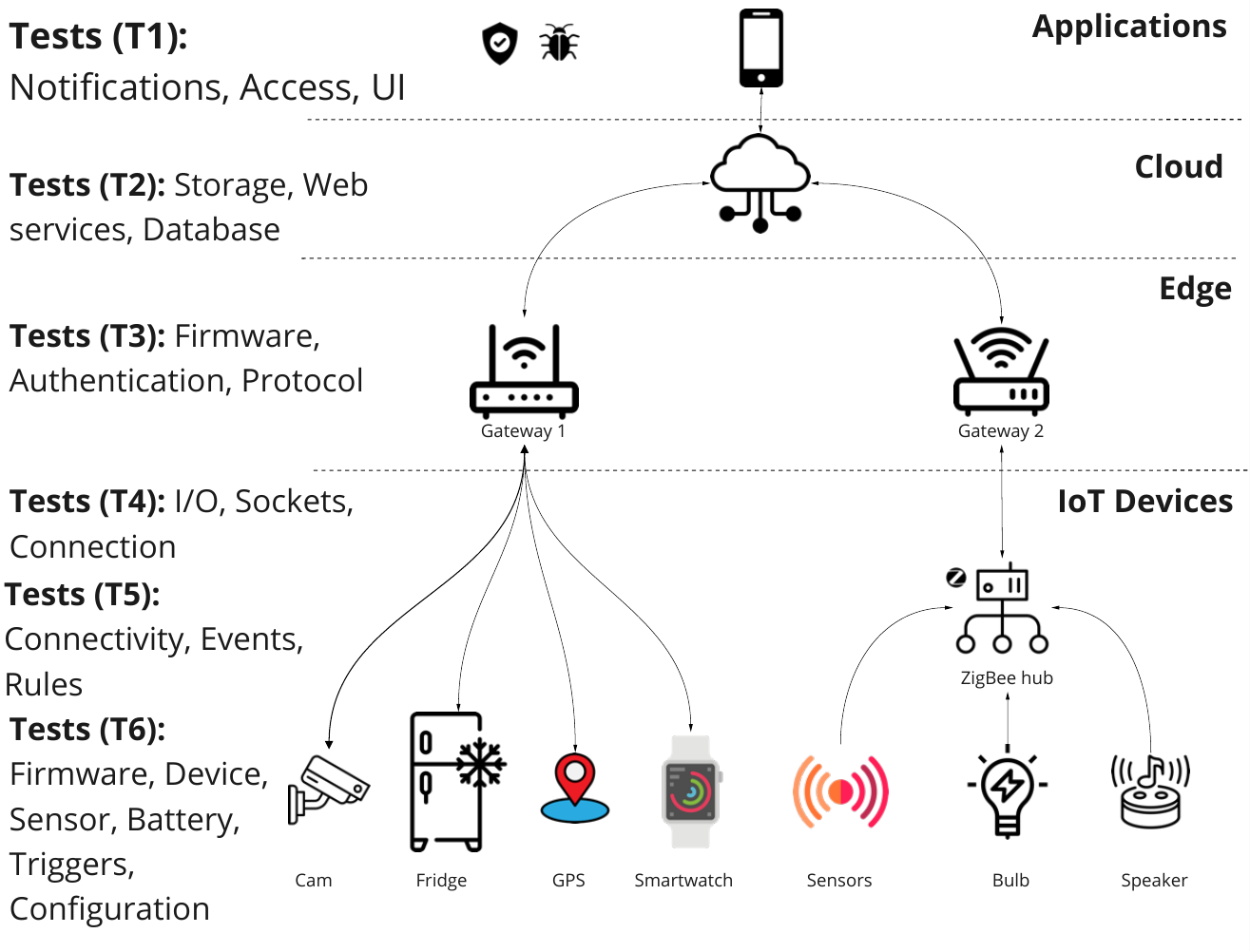}
  \caption{IoT platform layers and common tests per layer \cite{makhshari2021}}
  \label{fig:iot}
\end{figure}

\textbf{\iot platforms.} An \iot platform is a software suite that provides the framework and tools to facilitate the development, and management of applications and connected devices. Those platforms and devices constitute an ecosystem that automates places like homes. In general, any \iot platform follows the architecture and layers described in Fig. \ref{fig:iot}~\cite{makhshari2021}. \figref{fig:iot} depicts four layers (\ie \iot devices, edge, cloud, and applications). The \textit{\iot devices} layer involves smart programmable devices interacting with the physical world through embedded sensors and actuators. The \textit{edge} layer consists of gateway devices with fewer resource constraints, capable of locally handling telemetry data collection, processing, and routing. These gateways interpret diverse communication protocols like MQTT, CoAP, and HTTP, managing device-device and device-cloud interoperability. The \textit{cloud} layer comprises remote \iot cloud servers that accumulate and process telemetry data, communicating with heterogeneous \iot devices for remote control and monitoring. \iot cloud servers utilize a rule engine to enable users to write automation logic, defining interoperability behaviors within the \iot system. Finally, a user can utilize \textit{applications (apps)} to interact and control the \iot devices.

In our research, we evaluated open-source \iot platforms, ultimately choosing  \openhab~\cite{openhab} and \homeassistant because of their widespread adoption. %
Both platforms are vendor-agnostic supporting a large range of devices and network protocols and allowing to connect and interact with products from different manufacturers (\ie Amazon, Alexa, Nest, Hue, Homekit, etc). Both platforms allow users to create complex interaction rules (\ie turn off the lights at a given hour if the motion sensor is not activated) using a user interface. Both platforms are also actively supported by community and diverse developers offering a wide range of add-ons, integrations, and custom components.

\textbf{\iot testing.} Each layer within the \iot platform is susceptible to failure, and the impact can cascade to subsequent layers. Moreover, \iot devices, often deployed in unpredictable and challenging environments, may encounter extreme conditions. Potential tests are outlined in Fig.~\ref{fig:iot} are listed based on the most common test for software components \cite{linares-vasquez_enabling_2017}.

\textit{T1: Application Tests} - This layer focuses on user experience and interface testing to facilitate coordination and access to IoT devices. Common tests include UI/UX assessments, notification functionalities, and smartphone integration \cite{linares-vasquez_enabling_2017}.

\textit{T2: Cloud Tests} - IoT device data is transmitted to cloud-based platforms for storage, analysis, and accessibility. Tests in this layer typically cover storage integrity, database access, and web service functionalities.

\textit{T3: Communication Protocols} - Various communication protocols such as Wi-Fi, Bluetooth, and cellular networks enable IoT devices to connect to the internet and other devices. Tests here include router firmware evaluations, network authentication, and protocol compliance checks.

\textit{T4: Network Connectivity} - IoT devices require stable connections to broader networks through standard protocols. Tests focus on input/output data validation, socket connections, and data delivery reliability.

\textit{T5: Intermediate Hubs} - Intermediate hubs facilitate device control with specific requirements. Tests involve event validation, connectivity checks, and rules configuration, where rules define configurable routines for device actions.

\textit{T6: Firmware and Resource Optimization} - IoT devices utilize firmware to monitor, control, and optimize processes. Tests concentrate on optimizing limited resources such as battery and memory, alongside fundamental configuration and trigger event validations.

A common functional software artifact deployed on each device is the \addon. For instance, devices such as cameras, sensors, and speakers can be seamlessly integrated into the platform as \addons. Each \addon comprises multiple software \textit{components}, including but not limited to battery management, sensor control, event handling, data collection, and network connectivity. These \textit{components} serve as the fundamental \textit{building blocks across all \iot layers}. Consequently, components are omnipresent throughout the \iot platforms. Each platform can have collaborations with private companies such as Hue, Amazon Alexa, or Apple HomeKit.

\textbf{Focal Methods.} Our goal is to identify testing practices across widely-used open-source \iot platforms. One common approach for assessing test coverage involves running tests and determining which functional blocks and lines of code are executed. However, applying this method across diverse programming languages, add-ons, and core components— in \iot platforms like \openhab and \homeassistant — poses significant challenges. Executing test coverage in these environments requires configuring various files, setting up specific environments, and integrating multiple devices for testing purposes \cite{8919324,bosmans_testing_2019}.

Our investigation adopts an agnostic approach to identify tests within \iot add-ons and component code. To achieve this, we utilize the concept of focal methods to pinpoint functional and test methods. According to Tufano et al. \cite{tufano_methods2test_2022}, a focal method \textit{(fm)} refers to a method within a functional block that has defined parameters. Each \textit{fm} can be tested by zero or $n$ focal test methods. A focal test method \textit{(ftm)} is a functional code block specifically designed to test either a portion or the entirety of an \textit{fm}. The primary purpose of \textit{ftms} is to capture the intent and structure of human-written tests, ensuring traceability between tests and code under test.

Focal methods provide a mechanism for identifying both functional methods and their corresponding tests within a component. The concept of a \textit{fm} (\figref{fig:focalmethods}) is based on the principle that all code is organized into files, and these files follow a structured format dictated by the syntax of the programming language (PL) \cite{tufano_methods2test_2022}. In object-oriented PLs (e.g., Java), this structure is particularly clear, and five levels of context can be identified:

\textit{Focal Method (fm)}: this first level represents a set of public methods with bodies that contain the most critical information for generating test cases. These are the functional code blocks that can be executed and tested.

\textit{Class Context (fm + c)}: the second level adds the class name to the focal method, providing the necessary context about the class to which the method belongs.

\textit{Class Constructor}: includes the class constructor, which offers details on how to instantiate the class to enable testing.

\textit{Auxiliary Methods}:  comprise supporting methods (\eg getters and setters) that are required to invoke the focal method.

\textit{Class Attributes}: encompass the set of attributes at the class level, which contribute to the functionality of the class.

\begin{figure}[t]
  \centering
  \includegraphics[width=0.48\textwidth]{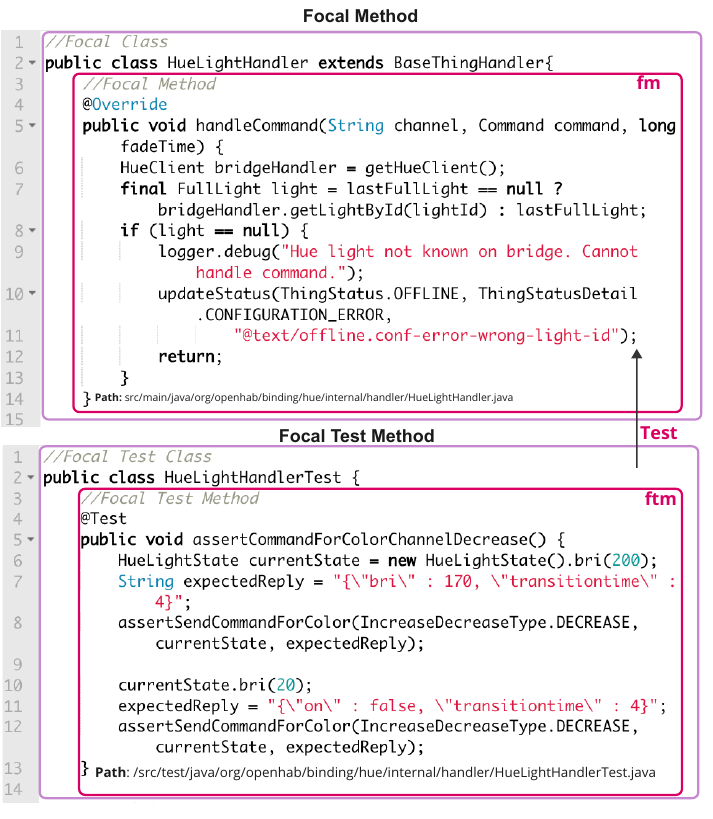}
  
  \caption{Focal method (fm) and focal test method (ftm) example.}
  \label{fig:focalmethods}
\end{figure}

\section{Research Questions}
\label{sec:rq}

In this research, we seek to learn more about \iot developers' perspectives regarding their test objective, common practices and techniques, test case design decisions, and challenges they face. By understanding what components are commonly tested, and the techniques used, we can not only gain insights into what areas might benefit from the development of new tools or techniques but also identify which parts of the system developers view as most critical or complex. This will guide the efficient allocation of testing resources and better focus future testing strategies, ultimately improving the reliability and performance of IoT applications. To evaluate how developers are currently testing \iot platforms, the practices they follow, and the challenges they face, we pose the following RQs:

\begin{enumerate}[label=\textbf{RQ$_{\arabic*}$}, ref=\textbf{RQ$_{\arabic*}$}, wide, labelindent=5pt]\setlength{\itemsep}{0.2em}
	
\item{\label{rq:common} \textit{What are the proportion of focal test methods for \addons and commonly components in open-source \iot platforms?}} 
\item{\label{rq:purpose}\textit{What is the purpose of current testing-related code and processes in open-source \iot platforms?}}
\item{\label{rq:design} \textit{How do developers design tests for \iot platforms?}}
\item{\label{rq:techniques}\textit{What tools, techniques, and processes do developers currently employ for testing \iot platforms?}}
\item{\label{rq:challenges}\textit{What challenges do developers face when testing \iot platforms?}}
\end{enumerate}

\section{IoT Platform Analysis}\label{platforms}

The following section describes the steps for collecting, curating, standardizing, and labeling each component from \openhab~\cite{openhab} and \homeassistant~\cite{homeassistant} as well as includes the results from our analysis.

\subsection{Methodology}

To answer our \ref{rq:common} we designed a semi-automated pipeline to collect and classify testing artifacts we found from \openhab and \homeassistant.

\begin{figure}[ht]
  \centering
  \includegraphics[width=0.48\textwidth]{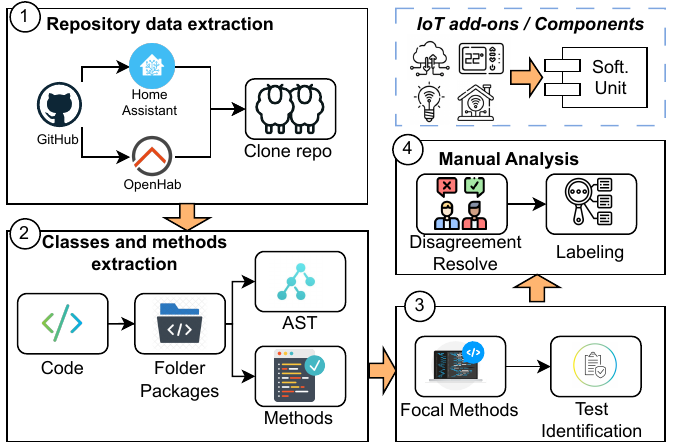}
  \caption{\homeassistant and \openhab platform analysis steps.}
  \label{fig:methodology}
\end{figure}

\subsubsection{Data extraction}
Our data is collected from the public \github repositories reported at \homeassistant and \openhab platforms (see Figure \ref{fig:methodology} \circled{1}). Each platform reports a set of projects depending on the core architecture, brands, or operative systems. Notably, each platform adopts a distinct code organization and architecture. For instance, \homeassistant,  designates a \texttt{\textit{test}} directory within its repository, exclusively housing test codes for each \addon. While \openhab developers maintain test codes for each \addon within their project folders. \openhab's repositories exhibit a variety of code languages, with Java being the most prevalent, accompanied by instances of Python, JavaScript, and Kotlin (see Table \ref{tab:test_files_investigation}). We noticed that \openhab lacks a standardized naming convention for Java classes, with names such as ``Test'', ``Stub'', or ``Mock'', adding a layer of complexity to the analysis.

\subsubsection{Automated test identification}
The step-\circled{2} includes the identification of the source folder, classes, and methods. We use regex expressions to isolate folders with functional code from tests. For instance, Java-based \addons, usually match the test folder and file name (\eg \texttt{src/test/java/FooTest.java}) to the corresponding functional code path (\eg \texttt{src/main/java/Foo.java}). \figref{fig:focalmethods} illustrates a linked \textit{ftm} with the correspondent \fm \texttt{HueLightHandler} class. That is not the case for different PLs like JavaScript or Python. In that case, we look for name matching, for example,  \texttt{x.python} is tested by \texttt{x\_test.python}. We also aim to identify classes and methods via abstract syntax tree (AST) with tree-sitter tool \cite{noauthor_tree-sitterintroduction_nodate}. The AST helps to confirm a valid code file and identifies structures, such as the methods, functions, and variables. 
\subsubsection{Test analysis}
\ref{rq:common} requires ascertaining the number of tests per component and their proportional relationship with the functional code. Therefore, we focused on identifying \fm and \ftm as described in \secref{sec:bakground} at step-\circled{3}. A \fm exclusively incorporates the signature, parameters, and body function. To identify a test we look not only to the test folder and files but also a \ftm inside each file (\ie method within a test class with the \textit{@Test} annotation) ( \figref{fig:methodology}). Once we identify the \textit{focal methods} and \textit{focal test methods} per file, we count them to obtain a proportion by file, component, \addon, and platform.

\subsubsection{Manual Analysis}

The fourth step-\circled{4} aims to validate the identification of focal methods and focal test methods. It allows us to confirm our proportion number to answer \ref{rq:common} and also enables the labeling process to answer \ref{rq:purpose}.

To inspect the  \openhab platform, we selected the top 36 \addons ranked by the highest count of \ftm and classes. Then we analyzed the top 10 test files with the most significant number of methods and classes. In this way, we ended up manually analyzing a total of 360 test files from a total of 5,597 test files.  Labeling is about assigning a common category to the given \ftm intention. In this step, two authors look through the \ftm to describe their intended purpose and functionalities. The authors scrutinize each test file to identify the components under examination and assign multiple labels accordingly. For instance, in files such as \texttt{test\_sensor.py}, developers frequently assess functionalities such as air quality, temperature, storage component setups, signal strength, network speed, voltage stability, and battery levels. These features are then generalized by labeling them as \texttt{`sensor', `storage',} and \texttt{`network'} (see Figure \ref{fig:label_distribution}).

The manual inspection procedure for \textit{Home Assistant} includes an additional step. We observed that \homeassistant repeats a set of filenames across multiple \addons. For instance, \texttt{test\_config.py} appears in 552 different \addons, and \texttt{test\_sensor.py} appears in 267 \addons.  The test files with the same name serve similar testing objectives over different \addons; For example, \texttt{test\_notify.py} tests various types of asynchronous notifications for both Google Mail and Slack components and \addons. Leveraging this fact, we narrowed our focus to the top 43 filenames, each appearing more than 10 times. For each of these 43 filenames, we randomly selected 10 instances and conducted a manual inspection of the source code. We selected the top filenames since they represent the biggest \iot systems and 10 is the average to frequency. This yielded a total of 430 manually inspected files for \homeassistant. The number of inspected files for \openhab and \homeassistant constitutes a representative sample, providing a confidence level (z-score) above 95\% (see Tab.~\ref{tab:test_files_investigation}).

Table \ref{tab:test_files_investigation} summarizes our test file selection process. The labeling process was executed in two phases. After each phase, the authors met to finalize the labels and resolve disagreements. This collaborative approach ensured alignment and prevented mislabeling or focus on incorrect label categories.

\subsection{Results from the Analysis}
In this section, we present our findings after applying our methodology for extracting both \textit{focal methods} (fm)  and \textit{focal test methods} (ftm) for \openhab and \homeassistant platforms. Table \ref{tab:test_files_investigation} outlines popular \addons for each platform ordered by their number of source files. Some \addons are well-known brands such as Alexa\cite{alexa}, Nest\cite{nest}, or Homekit\cite{homekit}. We calculate the test ratio as $ftm / (ftm + fm)$, with a threshold of 0.5 to flag low scores. While this threshold doesn't guarantee each \textit{fm} has a corresponding \textit{ftm}, it helps assess the \textit{ftm} distribution. We report the average and standard deviation.

\begin{table}[ht]
\centering
\caption{\small Most common \addons per platform and test ratio}

\scalebox{0.7}{%
\begin{tabular}{clccccc}
\hline
\multicolumn{7}{c}{\textbf{OpenHab common add-ons test ratio score}} \\ \hline
\textit{\textbf{Add-on}} &  & \textit{\textbf{Source files}} & \textit{\textbf{Test files}} & \textit{\textbf{fm}} & \textit{\textbf{ftm}} & \textit{\textbf{Test Ratio}} \\ \cline{1-1} \cline{3-7} 
tapocontrol &  & 8 & 8 & 399 & 35 & 0.08 \\
enigma2 &  & 6 & 6 & 100 & 109 & {\ul 0.52} \\
mielecloud &  & 66 & 65 & 766 & 632 & 0.45 \\
loxone &  & 32 & 31 & 274 & 196 & 0.42 \\
omnikinverter &  & 10 & 1 & 39 & 24 & 0.38 \\
wemo &  & 52 & 11 & 199 & 58 & 0.23 \\
\rowcolor[HTML]{EFEFEF} 
nest &  & 109 & 22 & 562 & 109 & 0.16 \\
\rowcolor[HTML]{EFEFEF} 
hue &  & 12 & 11 & 982 & 125 & 0.11 \\
irobot &  & 1 & 1 & 74 & 6 & 0.08 \\
\rowcolor[HTML]{EFEFEF} 
samsungtv &  & 32 & 32 & 218 & 0 & 0.00 \\
\multicolumn{1}{l}{} &  & \multicolumn{1}{l}{} & \multicolumn{1}{l}{} & \multicolumn{1}{l}{} & \multicolumn{1}{l}{} & \multicolumn{1}{l}{} \\
\multicolumn{1}{r}{\textit{Total}} &  & 7071 & 5597 & 76453 & 4585 & - \\
\multicolumn{1}{r}{\textit{avg{[}std{]}}} &  & - & 39.41{[}68.65{]} & 188.31{[}220.86{]} & 11.29{[}43.03{]} & 0.04{[}0.09{]} \\
\multicolumn{1}{r}{\textit{min - max}} &  & - & 1 - 632 &  & 0 - 635 & 0 - 0.60 \\ \hline
\multicolumn{7}{c}{\textbf{Home Assistant common add-ons test ratio score}} \\ \hline
\multicolumn{1}{l}{recorder} &  & 57 & 51 & 714 & 1114 & {\ul 0.61} \\
\rowcolor[HTML]{EFEFEF} 
\multicolumn{1}{l}{\cellcolor[HTML]{EFEFEF}tplink} &  & 23 & 27 & 108 & 156 & {\ul 0.59} \\
\multicolumn{1}{l}{template} &  & 26 & 34 & 304 & 331 & {\ul 0.52} \\
\multicolumn{1}{l}{hassio} &  & 24 & 20 & 235 & 247 & {\ul 0.51} \\
\multicolumn{1}{l}{unifiprotect} &  & 29 & 33 & 260 & 261 & 0.50 \\
\multicolumn{1}{l}{esphome} &  & 39 & 39 & 318 & 306 & 0.49 \\
\multicolumn{1}{l}{zha} &  & 36 & 33 & 348 & 332 & 0.49 \\
\multicolumn{1}{l}{matter} &  & 29 & 69 & 175 & 166 & 0.49 \\
\multicolumn{1}{l}{group} &  & 22 & 21 & 207 & 195 & 0.49 \\
\rowcolor[HTML]{EFEFEF} 
\multicolumn{1}{l}{\cellcolor[HTML]{EFEFEF}homekit} &  & 27 & 30 & 316 & 296 & 0.48 \\
\multicolumn{1}{l}{} &  & \multicolumn{1}{l}{} & \multicolumn{1}{l}{} & \multicolumn{1}{l}{} & \multicolumn{1}{l}{} & \multicolumn{1}{l}{} \\
\multicolumn{1}{r}{\textit{Total}} &  & 9176 & 7307 & 40480 & 32735 & - \\
\multicolumn{1}{r}{\textit{avg{[}std{]}}} &  & 9.80{[}5.92{]} & 7.81{[}9.16{]} & 43.24{[}60.84{]} & 34.97{[}79.47{]} & 0.42{[}0.16{]} \\
\multicolumn{1}{r}{\textit{min - max}} &  & 2 - 57 & 0 - 110 & 1 - 714 & 0 - 1654 & \multicolumn{1}{r}{0.02 - 0.88} \\ \hline
\end{tabular}
}
\vspace{0.1cm}
{\\ \footnotesize{* \addons with more than 20 source files for \homeassistant. Gray rows indicate well-known components. Underscore indicates above the threshold. Bottom the total, avg, min, and max number of elements across all \addons. }}
\vspace{-0.5cm}
\label{tab:test_files_investigation}
\end{table}

In our analysis of \openhab, we identified $406$ \addons with a total of $\approx7K$ source files. We observe a low test ratio averaging only $0.04$. That means that there are more functional methods than methods for testing them. Notably, some components like SamsungTV lack any identified \ftm in Java versions. We found a total of $\approx76K$\ \fm, but just identified $\approx4K$ \ftm. Additionally, our analysis extended to extra \openhab repositories not included as \addons in the main project. For example, we observe that Z-Wave and Alexa exhibit test ratios of $0.25$ and $0.2$, respectively.

For \homeassistant, we observe higher test ratios observing an average of $0.42$ but with a high variability of $0.16$. Therefore, we could observe some components with even more \textit{ftm} than \textit{fm} (\ie recorder) but some components with a low test ratio (\eg homekit). Interestingly, well-known brands like Alexa and google\_assistant both with $0.38$ of test ratio.  We also noticed a lower number of test classes. This is due to the lack of class definitions, however, we identified $\approx40K$ \fm and $\approx32K$ \ftm across the total of \addons. 

In our analysis for \homeassistant, we identified  that $937$ \addons which contained test codes, only $327$ of those have a test ratio above the threshold of $0.5$. We select the $0.5$ threshold as the upper values represent beyond the half proportion on tested code, however, practitioners can use higher thresholds. In other words, \homeassistant reports $65\%$ of \addons with a poor number of tested methods. The same analysis for \openhab reports that just 3 \addons achieve this threshold.

\finding{Only 3 out of 406  \openhab  \addons surpass our test ratio threshold of 0.5. \homeassistant has 327 \addons that have a better test over the 0.5 threshold but reports high variability with a standard deviation of 0.16.} \label{find:results1}

\begin{figure}[t]
  \centering
  \includegraphics[width=\linewidth]{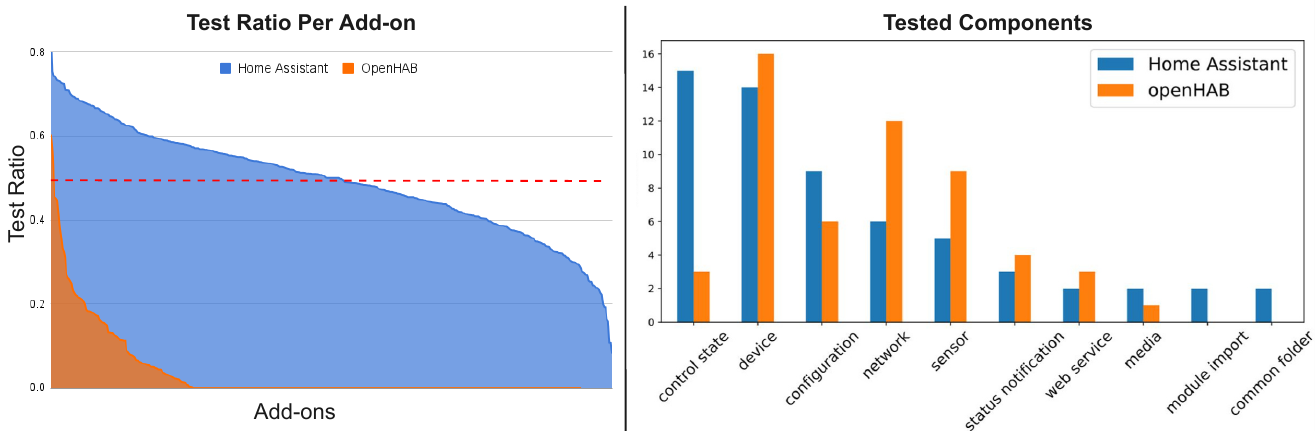}
  
  \caption{Left: Test ratio per \addon. Right: Top ten tested components for \homeassistant and \openhab}
  \label{fig:label_distribution}
\end{figure}

Figure~\ref{fig:label_distribution} on the left, depicts the test ratio among the \addons and the labeling components within \addons on the right. The test ratio distribution demonstrates a larger number of implemented \addons for \homeassistant and a stable number of \textit{ftm} above 0.2 but below 0.5. Nevertheless, \openhab test ratio is mostly below 0.2. 
\homeassistant's exclusive tested components like \texttt{`switch'}, \texttt{`button'}, \texttt{`cover'}, and \texttt{`trigger'} indicate the platform's concentrated efforts towards refining user-facing elements and interaction mechanisms. In contrast, \openhab's exclusive tested components such as \texttt{`rule'}, \texttt{`authentication'}, and \texttt{`status'} reflect the platform's strong emphasis on rule-based automation, security, and system organization. 

The components common to both \homeassistant and \openhab, such as \texttt{`scene'}, \texttt{`light'}, \texttt{`configuration'}, and \texttt{`sensor'}, signify functionalities fundamental to any smart home platform. These components represent core elements necessary for device management, configuration settings, and environmental sensing. \texttt{`Events'}, \texttt{`control state'}, and \texttt{`status notification'} highlight the focus on real-time updates and event-driven actions, ensuring users stay informed about their smart home ecosystem's status (see Figure \ref{fig:label_distribution} top). These components signify a common commitment between \homeassistant and \openhab to ensure the reliability, functionality, and interoperability of essential features within smart homes.

\finding{\homeassistant developers concentrated on enhancing user-facing elements and interactions, whereas \openhab developers prioritized rule-based automation. However, both platform developers focused on the core functionalities of smart home platforms.} \label{find:results2}

\begin{boxK}
\ref{rq:common} \fnumber{1}  and \fnumber{2} Demonstrate that \openhab has a poor test ratio and \homeassistant has $65\%$ \addons below the test ratio threshold. The most common tested components are devices, network, and sensor parameters for \openhab and control state, device, and configuration for \homeassistant. \homeassistant exclusively tests components such as switch, button, and cover, while \openhab reports rule-based, status, and ecosystem group tests.
\end{boxK}

To answer \ref{rq:purpose} we provided 12 common test types commonly employed in regular software testing scenarios, such as web solutions, services, and applications  (\tabref{tab:findings}). We identify several of these testing types in our analyzed platforms. The simplest and most prevalent test type is the unit test. While \homeassistant and \openhab report usability tests, we note that the complete implementation and results are not available in the repositories. Brands and communities potentially could employ an issue tracker tool independent of the \github repositories. As a result, we categorize usability and regression testing as ``Maybe'', indicating a potential implementation. Interestingly, we observe automation pipelines for \homeassistant, encompassing platform deployment, and some device configuration and testing. However, similar scripts or pipelines are not evident for \openhab. Table~\ref{tab:findings} also maps the tests to participants' preferences, which we describe in Section~\ref{sec:survey}.

\begin{table}[ht]
\centering
\caption{\small Test types and the observed type of test in repositories. Challenges reported on each type of test in the survey}
\label{tab:findings}

\scalebox{0.73}{%
\setlength{\tabcolsep}{5pt}

\centering
\begin{tabular}{llcclccc}
\hline
\multicolumn{1}{c}{\textbf{Test process}} &  & \multicolumn{2}{c}{\textbf{Platforms \ref{rq:purpose}}} &  & \multicolumn{3}{c}{\textbf{Purpose using testing techniques}} \\ \cline{1-1} \cline{3-4} \cline{6-8} 
\multicolumn{1}{c}{\textit{\textbf{Test Type}}} &  & \textit{\textbf{Home Assitant}} & \textit{\textbf{Openhab}} &  & \textit{\textbf{Automated}} & \textit{\textbf{Manual}} & \textit{\textbf{Testing IoT}} \\ \hline
Unit Testing &  & \cmark & \cmark &  & ND & ND & \xmark \\
Integration Testing &  & \xmark & \cmark &  & ND & ND & \xmark \\
Functional Testing &  & \xmark & \xmark &  & \xmark & \cmark & \xmark \\
End-to-End Testing &  & \xmark & \xmark &  & ND & \cmark & \xmark \\
Performance Testing &  & \xmark & \xmark &  & \cmark & ND & \cmark \\
Security Testing &  & \xmark & \xmark &  & ND & ND & \xmark \\
Usability Testing &  & Maybe & Maybe &  & \xmark & \cmark & \xmark \\
Regression Testing &  & Maybe & Maybe &  & \cmark & \xmark & \xmark \\
Accessibility Testing &  & \xmark & \xmark &  & \xmark & \cmark & \xmark \\
Compatibility Testing &  & \xmark & \xmark &  & \xmark & \xmark & \cmark \\
Database Testing &  & \cmark & \cmark &  & ND & ND & \xmark \\
API Testing &  & \cmark & \cmark &  & \cmark & ND & \xmark \\
CI/CD Testing &  & \cmark & \xmark &  & \cmark & \xmark & \xmark \\
Exploratory Testing &  & \xmark & \xmark &  & \xmark & \cmark & \xmark \\
User Acceptance &  & \xmark & \xmark &  & \xmark & \cmark & \xmark \\ \hline
\end{tabular}

}
\end{table}

\begin{boxK}
\ref{rq:purpose} Tests for API services and device-cloud communication insurance. Some scripts test databases and parameter configurations for multiple devices. Continuous Integration is also considered in the process.
\end{boxK}

\section{IoT Developers' Perspectives on Testing}\label{sec:survey}

The analysis of platform code in Section~\ref{platforms} develops a data-driven characterization of the extent of testing in smart home platforms. 
However, this characterization must also be complemented with an understanding of {\em why} the state of testing is as it is.
To this end, we present a survey-based study of developer perspectives on IoT testing, with the goal of understanding the key pain points experienced by developers, their priorities, and preferences that may affect how testing is carried out. 
This section describes the design of our survey, our coding and thematic analysis approach, and the key findings.

\begin{table}[ht]
\caption{\small Open-ended questions from survey}
\rowcolors{2}{}{isabelline}
\centering
\scalebox{0.83}{%
\setlength{\tabcolsep}{5pt} 

\begin{tabular}{cp{9cm}}
\hline
\textbf{ID}           & \multicolumn{1}{c}{\textbf{Question}}                    \\ \hline
\textit{\textbf{OQ1}} & What are the testing techniques that you employ and why? \\
\textit{\textbf{OQ2}} & What is your process for designing test cases for different types of products?    \\
\textit{\textbf{OQ3}} & What tools/APIs/frameworks do you use to support the testing of IoT products?     \\
\textit{\textbf{OQ4}} & How do you evaluate the effectiveness of your test cases/suites?                  \\
\textit{\textbf{OQ5}} & How do you typically resolve flaky tests?                \\
\textit{\textbf{OQ6}} & What are the main challenges that you have encountered when testing IoT products? \\
\textit{\textbf{OQ7}} & What would you improve in the current software testing process for IoT products?  \\
\textit{\textbf{OQ8}} & What would you improve in the current software testing or debugging tools?        \\
\textit{\textbf{OQ9}} & Have you ever faced any specific debugging-related challenges for IoT products?   \\ \hline
\end{tabular}
}
\label{tab:open_questions}
\end{table}

\subsection{Survey Design} 
Our survey consists of several questions on current testing practices and preferences, challenges the developers face, and the improvements they envision and organized as follows:

\begin{itemize}[wide, labelwidth=0pt, labelindent=0pt]
\item {\bf Demographic Information}: basic demographic information such as age, gender, and education.
\item {\bf Background Information}:includes employment status, years of general programming experience, IoT-related programming experience, and testing experience. We also asked how the participants learned about software testing. %
\item {\bf Testing Practices and Preferences}:  type of IoT product participants work on and the type of documentation they use for specifying IoT-related requirements for those products. We then asked about their preferred testing approaches and why they follow them. Furthermore, we asked them to provide information about their process of designing test cases and the tools they use to support the process. Next, we asked them about their evaluation process of those test cases. We also asked them if they had encountered any flaky tests during their testing process and how they resolved them. Finally, we asked whether they created test cases for reported bugs and the origins of those bugs. \item {\bf Challenges and Expectations}: After obtaining information on IoT developers' current testing practices, we asked them about the challenges they encountered when testing and debugging the products. Furthermore, we asked them what improvements they desired in the current testing and debugging process and tools. 
\end{itemize}
\begin{table}[ht]
\centering
    \caption{Demographic information over n=80 participants }
\scalebox{0.6}{%
\begin{tabular}{ccccccccccc} 
\hline
 & \multicolumn{4}{c|}{\textbf{Age}} & \multicolumn{6}{c}{\textbf{Education}} \\
 & \textit{\textbf{18-29}} & \textit{\textbf{30-39}} & \textit{\textbf{40-49}} & \multicolumn{1}{c|}{\textit{\textbf{50-64}}} & \textit{\textbf{High School}} & \textit{\textbf{Collage}} & \textit{\textbf{Vocational}} & \textit{\textbf{Bachelor}} & \textit{\textbf{Master}} & \textit{\textbf{Doctorate}} \\ \hline
\textbf{n} & 6 & 48 & 24 & \multicolumn{1}{c|}{2} & 3 & 12 & 14 & 45 & 5 & 1 \\
\textbf{\%} & 7.5 & 60 & 30 & \multicolumn{1}{c|}{2.5} & 3.75 & 15 & 17.5 & 56.25 & 6.25 & 1.25 \\ \hline
\multicolumn{1}{l}{} & \multicolumn{1}{l}{} & \multicolumn{1}{l}{} & \multicolumn{1}{l}{} & \multicolumn{1}{l}{} & \multicolumn{1}{l}{} & \multicolumn{1}{l}{} & \multicolumn{1}{l}{} & \multicolumn{1}{l}{} & \multicolumn{1}{l}{} & \multicolumn{1}{l}{} \\ \hline
 & \multicolumn{2}{c|}{\textbf{Gender}} & \multicolumn{4}{c|}{\textbf{Programming Experience}} & \multicolumn{4}{c}{\textbf{Testing Experience}} \\
 & \textit{\textbf{Male}} & \multicolumn{1}{c|}{\textit{\textbf{Female}}} & \textit{\textbf{0-3}} & \textit{\textbf{3-5}} & \textit{\textbf{5-10}} & \multicolumn{1}{c|}{\textit{\textbf{10+}}} & \textit{\textbf{0-3}} & \textit{\textbf{3-5}} & \textit{\textbf{5-10}} & \textit{\textbf{10+}} \\ \hline
\textbf{n} & 72 & \multicolumn{1}{c|}{8} & 5 & 7 & 21 & \multicolumn{1}{c|}{47} & 19 & 20 & 39 & 2 \\
\textbf{\%} & 90 & \multicolumn{1}{c|}{10} & 6.25 & 8.75 & 26.25 & \multicolumn{1}{c|}{58.75} & 23.75 & 25 & 48.75 & 2.5 \\ \hline
\end{tabular}
}
    \label{tab:demographic}

\end{table}

\subsubsection{Participant Recruitment}
We recruited participants from multiple IoT platform developers' community forums (\eg OpenHab community, Home Assistant community, SmartTthing community, and Google Nest community) by posting a flier. 
We received a total of 186 responses, of which we discarded 106 responses
due to (i) failed attention check questions, (ii) not finishing the survey, (iii) ambiguous responses, or (iv) duplicate responses. Finally, we obtained 80 valid responses (denoted as \pnumber{1}--\pnumber{80}), which we analyzed and presented our findings in Section~\ref{survey-results}. Our survey took an average of 15 minutes to complete, and we offered a $10$ USD Amazon Gift Card to each participant. Table~\ref{tab:demographic} provides demographic information about all of 80
participants. 
Most participants were male (90\%), all were at least 18 years old, and most were between 30 to 49 years of age (90\%).

\subsubsection{Ethical Consideration}
The study protocol was approved by our Institutional Review Board (IRB). 
Participants were informed about the study's goal before participating, and they willingly provided their consent to participate in the study and to disclose anonymized survey responses and quotes.

\subsection{Coding and Analysis}
We used descriptive statistical analysis to present the quantitative results. 
To analyze nine free-text questions, we used thematic analysis with an inductive coding approach~\cite{braun2021thematic}. 
Two out of three authors randomly selected a question and coded the data independently. 
After completing the coding, authors met and discussed any disparity in their codes and finalized the code after reaching a consensus. 
After all the responses were coded, all three authors discussed to extract themes or patterns in the answers. %

\subsection{Results from the Analysis of Survey Responses}
\label{survey-results}
In this section, we analyze the responses provided by the participants and answer 
\ref{rq:design}, \ref{rq:techniques}, and \ref{rq:challenges}.

\subsubsection{Test design and evaluation} 
OQ$_{1}$ and OQ$_{3}$ responses were used to answer~\ref{rq:design}. To design appropriate tests for their apps, developers start by learning the test requirements. Then they create a test plan by determining test goals, which may involve identifying corner cases, and in the process, building test scenarios. As \pnumber{63} states, \textit{``Once I have a clear understanding of the product, I create a test plan that outlines the scope of testing, including the devices, protocols, and communication methods involved in the IoT system.''} Some developers emphasized performance testing to assess speed and scalability. Security testing is also deemed crucial to developers, especially, as \pnumber{16} states, \textit{``...for products dealing with sensitive information.''}

Test effectiveness is evaluated by assessing its maintainability and execution time. Coverage-based evaluation is another popular assessment in which developers consider requirements and use-case coverage as well as code coverage and mutation analysis. Compliance with organizational standards is also considered to measure test case effectiveness.
\begin{boxK}
\ref{rq:design} Developers emphasize creating a comprehensive test plan after defining the product and testing scope, focusing on devices, protocols, and communication. Performance and security testing are crucial for sensitive data, with evaluation based on maintainability, execution time, coverage, and compliance.

\end{boxK}

\subsubsection {Current testing practices and preferences}
Table \ref{tab:findings} from Section~\ref{platforms} not only maps common testing techniques to our observations (of the presence/absence of the techniques) in our analysis of OpenHAB and HomeAssistant, but also maps them to participants' responses regarding why they use (or don't use) the specified techniques (\eg for manual analysis, automated analysis, or IoT testing in general). 
We observe that developers lean towards one technique over the other based on the nature of the test. For instance, developers opt for manual testing to ensure requirement compliance, evaluate interfaces, explore corner cases, and validate solutions. Conversely, they emphasize that automated testing is beneficial for evaluating performance, ensuring regression tests, and implementing continuous integration. For certain test types, we could not find any related answers; thus, we designated them as "Not Defined" (ND), signifying a lack of evidence regarding the use of automated or manual techniques, particularly for security evaluation or database testing.

Participants mention a set of tools oriented to the test and quality assurance such as Cucumber and Kibana, In addition to those tools, testing \iot protocols and connectivity associated with scripting are essential for device and network testing. We observe that just a few of them are dedicated to monitoring and data visualization.

Our analysis of the comments from participants leads to an understanding of their preferences. 
Particularly, we found that almost all of our participants expressed a preference for automated testing, followed by manual testing and semi-automated testing. 
The participants value the automated tests mainly because of their efficiency, speed, and coverage. The automated test enables cross-platform testing, continuous testing, and performance testing. There is a focus on identifying defects early in the development process to reduce costs and improve reliability, as \pnumber{46} states, \textit{``Automated testing helps catch bugs early in the development process, reducing the cost of fixing them later.''}. %

\finding{ Automated testing is preferred by most participants. There is a focus on identifying defects early in the development process to reduce costs and improve reliability. Consistency in results and repeatability are also highlighted, along with the importance of generating detailed logs and reports. } \label{find:results3}

Some participants prefer manual testing to test new features or achieve flexibility to requirements changes. 
Moreover, developers emphasized the need to develop a deep understanding of the system, guided by human intuition, \eg as \pnumber{4} states, \textit{``I want to see the nuts and bolts of how a thing works so I can better understand how to make it work the way I want.''}.
Participants also claimed that security issues are easily verified using manual testing, which is particularly interesting,  given that a recently study Ami et al.~\cite{ampn24} found that in general, developers prefer automated security tools for testing over manual analysis, given their rigor and ability to catch what developers miss. %
Finally, developers expressed that manual testing also helps evaluate user experience and accessibility.

\finding{Some developers recommend manual testing for its ability to bring in human intuition, adaptability, and a nuanced understanding of user experiences which can help test real-world scenarios.}\label{find:results4}

Some developers adopt a semi-automated approach to balance automation and manual testing, including human judgment and interactions for programs that cannot be automated. As \pnumber{64} explains: \textit{``I use semi-automated testing because it allows me to strike a balance between manual testing and automation. Some aspects of IoT devices and applications require human judgment and interaction that can't be fully automated.''}. Testers decide to use a hybrid between automated and manual testing mostly because of the flexibility in performing regression tests adding new features and testing real-world scenarios.

Finally, we find that developers resolve flaky tests mostly by updating and reviewing environment setups, analyzing logs, and optimizing test scripts. For resolving issues within the source codes, developers analyze test logs, review test scripts, and fix synchronization issues by adjusting wait times or implementing retry mechanisms.

\begin{boxK}
\ref{rq:techniques}
Based on \fnumber{3} and \fnumber{4}, Developers favor automated testing for early defect detection, cost reduction, regression testing, and managing changes, supported by monitoring and visualization tools for timely failure identification. However, manual testing remains essential for special cases and log analysis.
\end{boxK}

\subsubsection{Current challenges and future research scope} 
Our survey revealed several challenges developers face when testing IoT platforms and devices (\ref{rq:challenges}).
A major challenge expressed by several developers was about ensuring seamless connectivity and communication between devices and platforms as challenging, as there is no common platform.
Particularly, developers expressed how testing compatibility across a vast array of devices supported by various platforms is a time-consuming task. 

\finding{A primary challenge encountered by developers in testing \iot apps is validating compatibility across various devices and platforms.}\label{find:results5}

Similarly, ensuring the devices continue to work seamlessly with over-the-air updates, including both the software and firmware, is a major concern.  There are many \iot devices, such as, motion sensors, which run continuously. Delivering an update to these devices poses a significant challenge due to the potential for operational disruptions. As \pnumber{17} states, \textit{``Firmware over-the-air (FOTA) updates can be problematic, especially when dealing with a large number of devices. Debugging issues related to the update process, like interrupted downloads or failed installations, requires thorough testing.''}

\finding{Developers face difficulties in debugging issues when a firmware update disrupts the functionality of an \iot product.}\label{find:results6}

Moreover, performance testing is another challenge for devices that operate on low resources. As \pnumber{12} states, \textit{``Identifying why a device fails to wake up or operate as expected in low-power states can be a complex task.''} Sensor data inaccuracy also affects the debugging process when network interference or extreme weather conditions introduce noise in the data.

\finding{Developers also face challenges during testing or debugging in low-power mode.}\label{find:results7}

Developers perceive that the current test infrastructure lacks equipment and tools to simulate real-world scenarios, including extreme weather conditions or network interruptions. Scalability to handle larger IoT deployments and conduct performance testing is another major concern. Compatibility testing is mentioned for ensuring support over a wider range of devices. Strengthening security tests is also important to identify vulnerabilities and potential weaknesses. Some think that better documentation can help facilitate knowledge sharing and collaboration. 

Current tools can benefit greatly if they offer automated test reporting, visualization, and prioritization. As \pnumber{45} states, \textit{``Debugging tools should offer more comprehensive support for analyzing and visualizing complex data structures and their changes during runtime.''} 
Cross-platform testing needs improvement according to some developers. As \pnumber{50} states, \textit{``Improved cross-platform support in debugging tools would be valuable, allowing developers to debug code running on different operating systems seamlessly.''} Real-time collaboration in testing is another talking point among the developers as it can allow multiple team members to debug and troubleshoot issues simultaneously.

\finding{Performance, scalability, real-world scenarios, and real-time collaboration are the major concerns for developers facing \iot components debug and test.  }\label{find:results8}

The open-ended responses gathered from the survey help us unveil some crucial aspects of \iot app testing that require attention and consideration. 
Particularly, we found that only a few developers worry about lack of standards and documentation, relative to the issues highlighted in \fnumber{8}.

\finding{Few developers are concerned about poor documentation and lack of organizational standards and procedures for testing \iot platforms. }\label{find:results9}

Finally, the acknowledgment of compliance with organizational standards as a metric for measuring test case effectiveness underscores the importance of aligning testing procedures with established benchmarks. Likewise, the emphasis on security testing within intricate \iot environments resonates as a critical area, where vulnerabilities or misconfigurations might remain undetected until a security breach occurs. 

\finding{Security testing is a critical area where vulnerability detection becomes more difficult due to the diversity of devices and components and lack of \iot oriented testing tools.}\label{find:results10}

\begin{boxK}
\ref{rq:challenges} \iot development faces challenges like communication platform limits, knowledge gaps, and unclear user requirements due to poor documentation. Key issues include testing diverse devices, addressing security, optimizing power, ensuring real-time responsiveness, and managing updates and firmware. Persistent problems involve standardization, third-party integration, and testing across networks.

\end{boxK}

\section{Discussion}

While we managed to find answers to our five RQs, the overarching goal of our research is two-fold: to 1) learn about the general testing practices of IoT developers, and 2) Uncover the key gaps related to testing IoT platforms. 
Based on these goals, we present the following discussion topics.

\subsection{Primary Focus on Unit Testing}
Our analysis of  \openhab and \homeassistant demonstrates that the primary focus is on unit testing. This empirical observation also resonates with findings from our user survey, \ie a majority of the participants solely rely on automated testing (\fnumber{3}), and manual testing techniques such as user acceptance or exploratory testing are generally absent from practice, except in rare cases (\fnumber{4}). 

Particularly, User Acceptance Testing (UAT) offers evaluation of an application from the end-user perspective and validates the readiness of its deployment to the real-world environment \cite{wang24}. During our survey, \iot developers did not emphasize on this type of testing. The reason for not employing UAT is because of its dependence on manual effort and its time consuming nature. Although recent works have focused on automating this process using large language models \cite{wang24}. Exploratory testing can help in the continuous integration and delivery pipeline for a large-scale software system\cite{maartensson2022chapter}, which was one of the more common concerns among the participants. 
While there is recent work on performing exploratory testing using static analysis \cite{doyle23} or in a gamified way \cite{coppola24}, additional research is necessary to evaluate the usability and efficiency of such techniques in the IoT context.

Similarly, we observe no integration testing in our analysis of \openhab and \homeassistant.
One explanation for this observation could be the sheer difficulty of testing across various diverse platforms and devices, as perceived by developers in our study (\fnumber{5}), particularly given concerns regarding the unpredictable impact of firmware/software updates (\fnumber{6}).

\subsection{Compatibility Testing and Future Solutions}
Several participants talk about the importance of compatibility testing in the context of \iot apps and integrations (\fnumber{5}), which is challenging due to the general fragmentation in the smart home landscape, in terms of platforms, communication protocols, networking standards, operating systems, and types of devices and sensors. 
Ensuring compatibility among these diverse technologies is crucial for IoT platforms.  
The absence of robust compatibility testing across all technologies may lead to post-release issues, ranging from user inconvenience to severe security vulnerabilities. This is a timely and critical challenge, as developers invest substantial time in repetitive tests across varied technologies, dealing with debugging issues that arise post-deployment.

Given the general lack of integration testing and the concern among developers regarding compatibility and testing in the fragmented IoT space, the integration of simulation environments into the testing apparatus offers a promising direction for future research. 
Simulating real-world scenarios involving varied device types and communication protocols would allow developers to validate changes with increased robustness. Furthermore, advancements in \iot interoperability frameworks, automation in testing procedures, and the potential infusion of machine learning or AI-driven testing solutions hold promise in mitigating compatibility challenges.

\subsection{Performance Testing and Scalability Challenges}
Managing scalability and performance in \iot software emerges as a complex task, particularly in large-scale deployments (\fnumber{8}). As the \iot platforms expands, devices get interconnected with a multitude of other devices and data points, which makes testing of these devices and apps more challenging. These challenges are crucial as they directly impact system reliability, efficiency, and user experience. This issue is inherently tied to \iot due to its vast array of user base, and continuous data exchange occurring among the devices. 

Upgrading the testing infrastructure to accommodate scalability testing at various levels of deployment can be a potential research direction. 
Tools enabling continuous integration to identify and resolve performance bottlenecks in real-time can help in mitigating this issue. There are existing tools available that address performance and scalability testing in various domains, such as Apache JMeter~\cite{halili2008apache}, Gatling~\cite{gatling}, Taurus~\cite{taurus}. However, their applicability in \iot scenarios may be limited due to challenges in simulating complex \iot environments and diverse communication protocols. These unique, IoT-specific challenges demand more specialized testing solutions.

\subsection{Improvements in IoT Testing Compliance}
The responses from \iot developers regarding organizational standards in testing (\fnumber{9}) reveal a significant emphasis on adhering to company policies, industry standards, and regulatory requirements. 
Testing for compliance with industry-specific regulations, such as medical device or automotive safety standards, or privacy regulations such as GDPR~\cite{gdpr} and CPRA~\cite{CPRA} in the context of smart homes, adds complexity to the testing process. 
Addressing these challenges requires an approach involving robust compliance testing methodologies with continuous monitoring. 
The \iot industry can benefit greatly from tools that can analyze historical compliance data, regulatory changes, and industry-specific standards to detect potential areas of non-compliance.

\section{Threats to Validity}\label{sec:threats_validity}
Threats to \textbf{construct validity} include concerns over the test ratio in automated analysis compared to test coverage, due to the complexity of components and environments. Traceability issues on some platforms obscure links between test and functional code. To address this, we combined quantitative analysis with manual inspection by two authors, providing insights into developers' testing intentions, supported by survey responses.

Threats to \textbf{internal validity} refer to the representativeness of the randomly selected tests at the labeling process and reported components.  
We sorted the \addons and components to identify the largest number of focal methods and focal test methods. We calculate the z-score on our test file sampling and manual inspection to achieve the 95\% confidence.

Another potential internal validity of our study lies in the subjectivity inherent in the manual analysis of test codes. To address this, we adopt a paired analysis approach, where analyses are conducted collaboratively.  This helps us minimize individual biases. To avoid misdirection while individual coding, we conduct our analysis in multiple phases. We observed a substantial decrease in disagreement rates as we progressed through the latter phases of analysis.

\section{Related Work}\label{sec:related_work}

While studies have examined testing practices for industrial software \cite{hynninen18,kochhar_practitioners_2019} and mobile apps \cite{vasquez17,Wang2019LookingFT,linares-vasquez_enabling_2017}, this research focuses on the challenges of software testing in \iot. It highlights persistent bugs despite significant time, resources, and testing efforts, as well as developers' practices and perceptions of test case design, automation, and quality metrics.

Research on \iot platforms has explored tools for bug detection \cite{9306923,bosmans_testing_2019,9014711, zhang_trace2tap_2020,brackenbury_how_2019} and highlighted the complexity of testing such systems \cite{8919324, REGGIO2020100313}. Bures \etal \cite{10.1007/978-3-030-58768-0_6} emphasize the need for specialized testing methods tailored to \iot. Surveys focus on bug detection, security failures \cite{9402092,electronics11091502,9878283,manandhar_helion_2019,jin_understanding_2022,AHANGER2022108771,ami_why_2022,makhshari2021}, and future directions. Zhu \etal \cite{zhu_survey_2022} predict trends toward intelligent, large-scale testing, emphasizing big data, cloud computing, and AI. Increased adoption of Smart Home solutions has led to security evaluations, such as Google's Nest and Philips Hue platforms, revealing key vulnerabilities and potential misuse \cite{kafle_study_2019,nest}.

Previous research focuses individually on tools and methods, introducing challenges on specific topics like cloud computing or some security concerns, this study aims to acquire a deeper understanding of the testing methodologies currently employed by \iot developers. We seek to grasp the big picture about how developers are testing \iot platforms and the rationale behind their choices and collect insights regarding potential areas for enhancing future testing practices in \iot.

\section{Conclusions}\label{sec:conclusions}

Our examination of the landscape of software testing within \iot platforms derived substantive insights, with 10 key findings from both a mining-based study and a survey with developers. 

We took a closer look at two specific platforms – \homeassistant and \openhab. 
We find notable evidence signaling the difficulty that developers face with testing IoT platforms, with the majority of \addons and integration apps of both platforms falling short of the 50\% test coverage threshold.
On average, only 5\% \addons contains any test methods for \openhab and well-known brand like Amazon Alexa exhibits a maximum test ratio of 59\%.

A majority of our survey participants prefer automated testing, to try to catch problems early in the development process to save time and make things more reliable. They stress the need for consistent, repeatable tests and detailed logs. But interestingly, some developers still prefer manual testing as it facilitates human intuition and adaptability for real-world situations.

Finally, our research identifies challenges in fixing problems caused by software updates, cross-platform testing, and dealing with issues in low-power devices -- among others. In summary, our study sheds light on the testing practices, tools, perceptions, and challenges for IoT platforms, illustrating promising pathways for future research to improve testing for this rapidly growing domain.

\ifCLASSOPTIONcompsoc
\else
\fi

\ifCLASSOPTIONcaptionsoff
  \newpage
\fi

\bibliographystyle{IEEEtran}
\bibliography{utils/iot}

\end{document}